\documentclass[letterpaper,twocolumn,prl,aps,showspace,superscriptaddress,floatfix,amsmath,amssymb]{revtex4}
\usepackage[latin1]{inputenc}
\usepackage{bm}
\usepackage{multirow,amssymb,amsbsy,amsmath}
\usepackage{graphicx}
\usepackage{verbatim}
\makeatletter
\usepackage{pifont}
\makeatother

\usepackage{graphicx}
\usepackage{dcolumn}
\usepackage{bm}
\usepackage{ifthen}
\usepackage{booktabs}
\usepackage{nicefrac}
\usepackage{float}

\begin{document}

\title{Reconstruction of a Photonic Qubit State with Reinforcement Learning}
\author{Shang Yu}
\affiliation{CAS Key Laboratory of Quantum Information, University of Science and Technology of China, Hefei 230026, People's Republic of China}
\affiliation{CAS Center For Excellence in Quantum Information and Quantum Physics, University of Science and Technology of China, Hefei 230026, People's Republic of China}
\author{F. Albarr\'an-Arriagada}
\affiliation{Departamento de F\'isica, Universidad de Santiago de Chile (USACH), Avenida Ecuador 3493, 9170124, Santiago, Chile}
\affiliation{Center for the Development of Nanoscience and Nanotechnology 9170124, Estaci\'on Central, Santiago, Chile}
\author{J. C. Retamal}
\affiliation{Departamento de F\'isica, Universidad de Santiago de Chile (USACH), Avenida Ecuador 3493, 9170124, Santiago, Chile}
\affiliation{Center for the Development of Nanoscience and Nanotechnology 9170124, Estaci\'on Central, Santiago, Chile}
\author{Yi-Tao Wang}
\affiliation{CAS Key Laboratory of Quantum Information, University of Science and Technology of China, Hefei 230026, People's Republic of China}
\affiliation{CAS Center For Excellence in Quantum Information and Quantum Physics, University of Science and Technology of China, Hefei 230026, People's Republic of China}
\author{Wei Liu}
\affiliation{CAS Key Laboratory of Quantum Information, University of Science and Technology of China, Hefei 230026, People's Republic of China}
\affiliation{CAS Center For Excellence in Quantum Information and Quantum Physics, University of Science and Technology of China, Hefei 230026, People's Republic of China}
\author{Zhi-Jin Ke}
\affiliation{CAS Key Laboratory of Quantum Information, University of Science and Technology of China, Hefei 230026, People's Republic of China}
\affiliation{CAS Center For Excellence in Quantum Information and Quantum Physics, University of Science and Technology of China, Hefei 230026, People's Republic of China}
\author{Yu Meng}
\affiliation{CAS Key Laboratory of Quantum Information, University of Science and Technology of China, Hefei 230026, People's Republic of China}
\affiliation{CAS Center For Excellence in Quantum Information and Quantum Physics, University of Science and Technology of China, Hefei 230026, People's Republic of China}
\author{Zhi-Peng Li}
\affiliation{CAS Key Laboratory of Quantum Information, University of Science and Technology of China, Hefei 230026, People's Republic of China}
\affiliation{CAS Center For Excellence in Quantum Information and Quantum Physics, University of Science and Technology of China, Hefei 230026, People's Republic of China}
\author{Jian-Shun Tang}
\email{tjs@ustc.edu.cn}
\affiliation{CAS Key Laboratory of Quantum Information, University of Science and Technology of China, Hefei 230026, People's Republic of China}
\affiliation{CAS Center For Excellence in Quantum Information and Quantum Physics, University of Science and Technology of China, Hefei 230026, People's Republic of China}
\author{E. Solano}
\affiliation{Department of Physical Chemistry, University of the Basque Country UPV/EHU, Apartado 644, 48080 Bilbao, Spain}
\affiliation{IKERBASQUE, Basque Foundation for Science, Maria Diaz de Haro 3, 48013 Bilbao, Spain}
\affiliation{Department of Physics, Shanghai University, 200444 Shanghai, China}
\author{L. Lamata}
\affiliation{Department of Physical Chemistry, University of the Basque Country UPV/EHU, Apartado 644, 48080 Bilbao, Spain}
\author{Chuan-Feng Li}
\email{cfli@ustc.edu.cn}
\affiliation{CAS Key Laboratory of Quantum Information, University of Science and Technology of China, Hefei 230026, People's Republic of China}
\affiliation{CAS Center For Excellence in Quantum Information and Quantum Physics, University of Science and Technology of China, Hefei 230026, People's Republic of China}
\author{Guang-Can Guo}
\affiliation{CAS Key Laboratory of Quantum Information, University of Science and Technology of China, Hefei 230026, People's Republic of China}
\affiliation{CAS Center For Excellence in Quantum Information and Quantum Physics, University of Science and Technology of China, Hefei 230026, People's Republic of China}

\date{\today }
\begin{abstract}
We perform an experiment to reconstruct an unknown photonic quantum state with a limited amount of copies. We employ a semi-quantum reinforcement learning approach to adapt one qubit state, an ``agent'', to an unknown quantum state, an ``environment'', by successive single-shot measurements and feedback, in order to achieve maximum overlap. Our experimental learning device, composed of a quantum photonics setup, can adjust the corresponding parameters to rotate the agent system based on the measurement outcomes ``0'' or ``1'' on the environment (i.e., reward/punishment signals). The results show that, when assisted by such a quantum machine learning technique, fidelities of the deterministic single-photon agent states can achieve over $88\%$ under a proper reward/punishment ratio within 50 iterations. This protocol offers a tool for reconstructing an unknown quantum state when only limited copies are provided, and can also be extended to high dimensions, multipartite, and mixed quantum state scenarios.
\end{abstract}

\maketitle
\bibliographystyle{prsty}
\emph{\bf{Introduction.}}\textemdash Extracting information from an unknown quantum state is an important task in quantum information. For the most general way, the quantum state tomography has to measure the averages of a set of observables for reconstructing the density matrix~\cite{James2001}. This method requires enough number of copies of the target state and will become unfeasible when the target system is large. Thereafter, many new approaches are proposed for reconstructing the unknown quantum state, such as the efficient tomography, which only requires unitary operations or local measurements~\cite{Cramer2010}, and the recent neural network tomography, which is based on restricted Boltzmann machines (RBM)~\cite{Torlai2018N,Torlai2018}. In addition, inspired by fast developments in machine learning techniques~\cite{Chouard2015,Stajic2015,Silver2016,RussellNorvig}, here we want to explore a semiautonomous strategy in experiments that can acquire information from another quantum system and adapt to it without external intervention.

Machine learning~\cite{MehtaReviewML}, as a subtopic within AI realm, has already become a powerful tool for data mining, pattern recognition, among others. Meanwhile, there are many recent works combining ML techniques with quantum information tools~\cite{Biamonte2017,DunjkoReview18,PetruccioneReview,Sasaki2001,Dong2008,Harrow2009,Bisio2010,Wiebe2012,Lloyd2014,Paparo2014,UnaiQSL,Faccin2014,Cai2015,Li2015,Dunjko2016,Sentis2016,Carleo2017,Wang2017,Deng2017,Sentis2017,Torlai2018N,Yu2018,Gao2018,Torlai2018}. These include expressing and witnessing quantum entanglement by artificial neural networks (ANN)~\cite{Deng2017,Gao2018}, analyzing and restructuring a quantum state by restricted Boltzmann machines~(RBM)~\cite{Torlai2018N,Torlai2018,Carleo2017}, as well as detecting quantum change points, and learning Hamiltonians by Bayesian inference~\cite{Sentis2016,Sentis2017,Yu2018,Wang2017}. Meanwhile, many quantum ML algorithms have already been applied into different experimental systems, such as photonics~\cite{Cai2015} and nuclear magnetic resonance (NMR)~\cite{Li2015} systems. Besides the above tasks, here, we focus on the topic of reinforcement learning (RL), which its quantum versions have been recently proposed~\cite{Dunjko2016,Dong2008,Paparo2014,Lamata17,Cardenas18,Arriagada2018}, being inspired on RL algorithms existing since the beginning of AI. In this context, a quantum system, named ``agent'', can learn how to behave correctly through interactions with a quantum ``environment'' state, and the ``reinforcement'' signals---rewards or punishments~\cite{Dunjko2016}, obtained from interactions between both.

In our quantum optical experiments, we explore whether a quantum state adapts onto another unknown quantum state via limited single-shot measurements, where only one photon is measured in each iteration, such that the ``reinforcement'' signal can be detected. The landmark principle that a single unknown quantum state cannot be cloned~\cite{Wootters1982}, prevents one from exactly copying a quantum state in a single-shot. However, here, we are interested in how to obtain the largest amount of information from the unknown state with minimal resources, i.e., with minimal number of identical copies, assisted by the RL algorithm.

In this work, we employ the semi-quantum reinforcement learning (sQRL) protocol~\cite{QCRL} introduced in Ref.~\cite{Arriagada2018} to reconstruct an unknown quantum state without performing state tomography. In order to realize the single-shot measurement in this learning task, we build a pseudo on-demand single photon source by post-selection method and a chopper~\cite{Yu2018}. Meanwhile, we also employ a register qubit to interact with the environment system via a Controlled-NOT gate and avoid to detect the environment qubit directly. When the ``reinforcement" signal is detected from the register system, we use it to determine the reward or punishment and calculate the unitary operation for the next learning step. In our experiment, we employ 50 iterations, namely, only consume 50 copies of photons of the environment system, and find the fidelities of the agent systems can reach over $88\%$ under a proper reward/punishment ratio.

\emph{\bf{Semi-quantum reinforcement learning algorithm.}}\textemdash In this section, we review the semi-quantum reinforcement learning algorithm proposed in Ref.~\cite{Arriagada2018}. In this work, we mainly focus on the case of reconstructing a qubit state with minimal resources. Our quantum learning algorithm involves three systems: the environment system ($E$), the register system ($R$), and the agent system ($A$), as shown in Fig. 1(a). We assume that the environment system (learning target) can be an arbitrary single-qubit state
\begin{eqnarray}
|E\rangle=\cos(\theta_{E}/2)|0\rangle+e^{i\phi_{E}}\sin(\theta_{E}/2)|1\rangle,\label{InitialEstate}
\end{eqnarray}
while agent and register qubits are both initialized in $|0\rangle$.

First, we interact the register with the environment system. This step is not necessary if one can measure the environment system directly. Here, we present a more robust method in which the measurements are acted on the register qubit, such that this can be achieved irrespective of whether the environment system is measurable or not. To realize this interaction, we perform a controlled-NOT (CNOT) gate with $E$ as control and $R$ as target, which produces the state
\begin{eqnarray}
\vert \psi \rangle& = &U^{\text{CNOT}}_{E,R}|E\rangle|0\rangle_{R} \nonumber \\
& = &\cos(\frac{\theta_{E}}{2})|0\rangle_{E}|0\rangle_{R}+e^{i\phi_{E}}\sin(\frac{\theta_{E}}{2})|1\rangle_{E}|1\rangle_{R}.
\end{eqnarray}

Then, we extract the information from the register qubit by measuring it on the basis $\{|0\rangle, |1\rangle\}$. If we detect the signal in the $|0\rangle$ state (the outcome $m=0$), it indicates that the agent state is similar to the environment state with probability $p=\cos^{2}(\theta_{E}/2)$. On the contrary, if the signal is detected in the $|1\rangle$ state (the outcome $m=1$), it means the agent state is opposed to the environment state with probability $p=\sin^{2}(\theta_{E}/2)$.

\begin{figure}[tbph]
\centering
\includegraphics[width=0.45\textwidth]{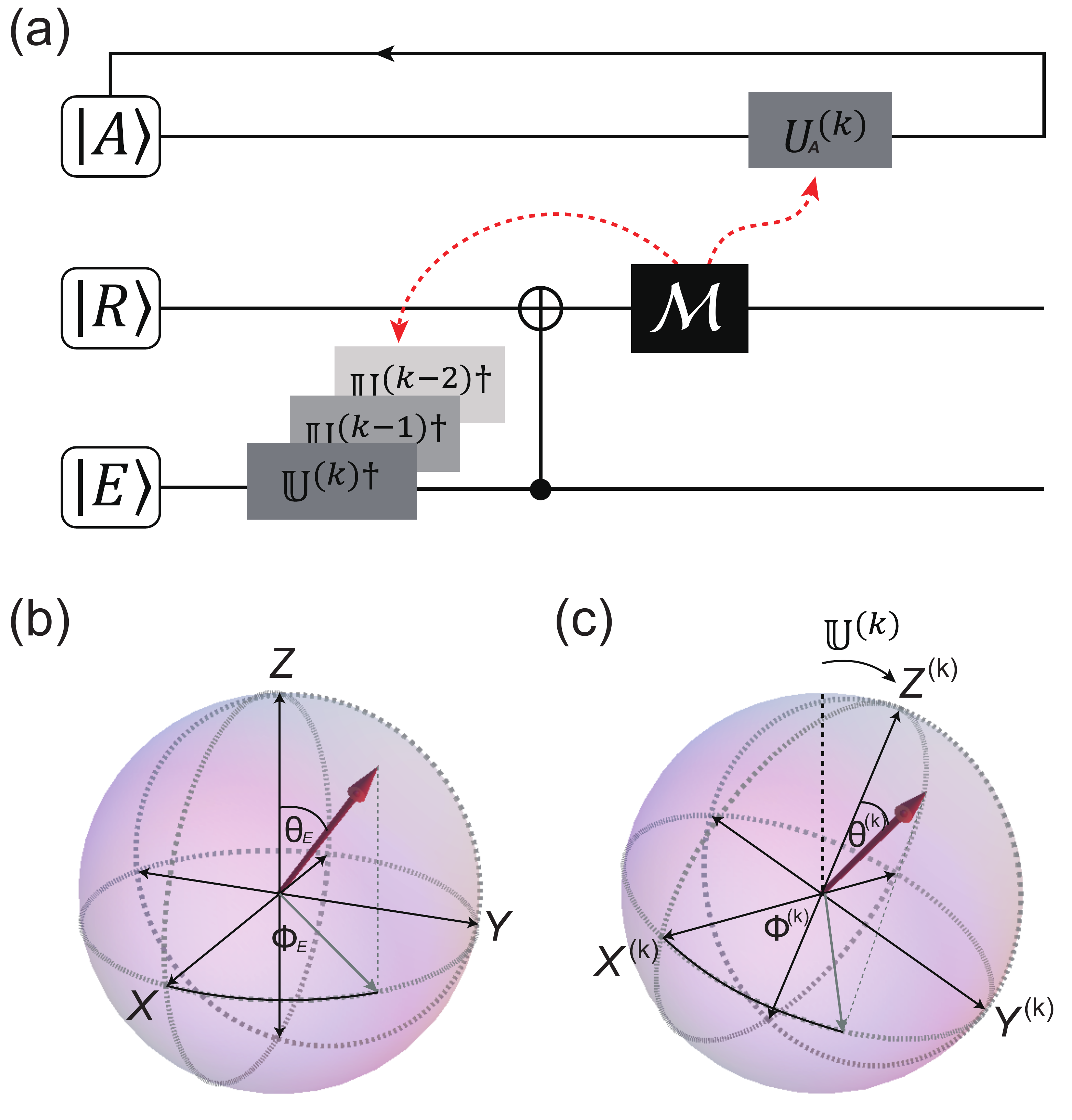}
\caption{(a) Quantum circuit diagram for the semi-quantum reinforcement learning protocol. The register qubit first interacts with the environment, and then is detected by a single-shot measurement. The measurement outcomes ``0'' or ``1'' which decide the rewards and punishments are input to a computer (not shown in the figure), and control the learning devices (unitary operations) by classical communications, denoted by red dashed lines. (b) Environment state at the initial iteration, shown as the red arrow. (c) Environment state at the $k$th iteration, in which the reference axes are rotated by previous iterations.}
\end{figure}

According to the information above, we perform a corresponding action on the agent qubit. When the outcome is 0, we just do nothing, i.e., $U_{A}=\mathbb{I}$. However, when $m=1$, we perform a partially-random unitary operation $U_{A}=\bar{U}(\phi)\bar{U}(\theta)$ on the agent qubit. At $i$th iteration, $\bar{U}(\phi)=e^{-iS^{(i)}_{z}\phi^{(i)}}$ and $\bar{U}(\theta)=e^{-iS^{(i)}_{x}\theta^{(i)}}$ present the phase and amplitude operations along the direction of the spin of the agent. Here, $S_{z}^{(i)}$ and $S_{x}^{(i)}$ are the rotated Pauli matrix at $i$th iteration, and $\phi$, $\theta$ are the random angles with a range $\Delta$, which can be written as $\phi,\theta\in[-\Delta/2,\Delta/2]$.
The angle range is modified by a reward function, which is decided by the outcome of the last step: $\Delta^{(i)}=[(1-m^{(i-1)})\epsilon+m^{(i-1)}1/\epsilon]\Delta^{(i-1)}$. Here, $\epsilon$ ($0<\epsilon<1$) controls the reward and punishment ratios, i.e., the value of $\Delta$ will be reduced when state $|0\rangle$ is detected, and vice versa. In this step, the partially-random unitary operation becomes $\mathbb{U}^{(i)}=U_{A}^{(i)}\mathbb{U}^{(i-1)}$ ($\mathbb{U}^{(1)}=\mathbb{I}$). This can be regarded as performing a unitary operation $U_{A}^{(i)}$ along the reference axes obtained at $i-1$ iteration, shown in Fig. 1 (b) and (c).

The action of the operator $U^{(k)}_A\mathbb{U}^{(k-1)}$ over $A$ is equivalent to the action of the operator $\mathbb{U}^{(k)\dagger}$ over $E$, which changes the basis of the environment in order to perform the measurement process in the logical basis $0/1$.

\emph{\bf{Experimental setup.}}\textemdash The experimental setup is shown in Fig. 2, which can be recognized as five parts: the generation of photon pairs by periodically poled KTP (PPKTP) crystal through the spontaneous parametric down-conversion (SPDC) process; the environment system that can provide many copies of the unknown quantum state; the register system which can interact with the environment; and the agent, which can generate deterministic single-photons and be polarized at $|H\rangle$ ($H$ and $V$ represent the horizontal and vertical polarizations, respectively), is equipped with a learning device that can adjust the polarization component and the relative phase of the photon. The PPKTP crystal is pumped by a 404 nm laser with horizontal polarization, and a pair of photons at 808nm can be generated with horizontal and vertical polarization, respectively. A polarizing beam splitter (PBS) thereafter divides these two photons into different paths: one of them is sent to the environment system, and the other one to the register system. In the environment system, the photons are prepared in an unknown state in a black box. Then, a unitary operation, which is constituted by half-wave plate (HWP) and quarter-wave plate (QWP) and controlled by the measurement results adjusts the reference axis~\cite{Yu2017}. The register qubit is prepared in state $|H\rangle$ and interacts with the environment system by a CNOT gate.

\begin{figure}[tbph]
\centering
\includegraphics[width=0.48\textwidth]{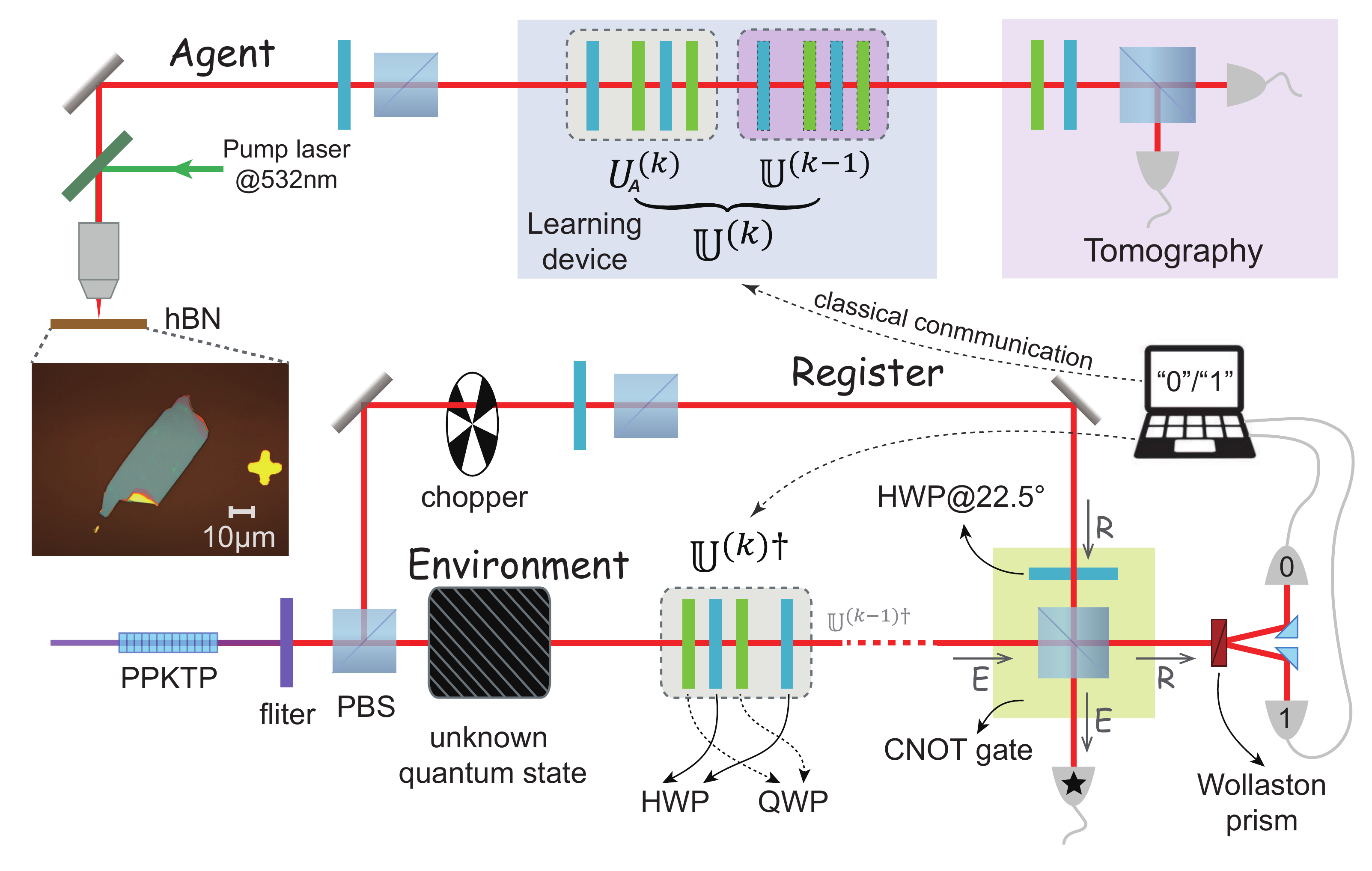}
\caption{Experimental setup. The photon pairs are generated from PPKTP by SPDC process, and work as environment system and register system, respectively. On the environment route, photons are randomly prepared and go through a unitary gate which is determined by the last iteration. On the register route, the chopper works as a switch for creating a pseudo-on-demand single photon source assisted by postselection. Then, the photons are prepared at $|H\rangle$ and interact with the environment by a CNOT gate. After the interaction, we measure the register qubit and obtain the outcome ``0'' or ``1'' (If we find the first event when chopper is switched on is the coincidence of ``0'' detector and ``$\bigstar$'' detector, we deem that the outcome at this iteration is ``0''). According to the outcome in the current iteration, we perform the action on the agent system as well as the environment system.}
\end{figure}

Considering the $k$th iteration, first we prepare the photons in the environment system at an unknown state, as shown in Eq.~(\ref{InitialEstate}) in a black box. This time the unitary operation $\mathbb{U}^{k-1}$ is settled by $(k-1)$th iteration. Then, the register qubit, which is prepared at $|H\rangle$, interacts with the environment system by a CNOT gate~\cite{He2016,Lu2007} (more details about the CNOT gate are shown in the Appendix A). After this, we extract the information by measuring the register qubit at $\{|H\rangle,|V\rangle\}$ basis via a Wollaston prism, and we can obtain the outcomes $m=0$ or $m=1$. According to the measurement outcomes, we calculate the unitary operation $U_{A}^{(k)}$, including $\theta$ and $\phi$, on the classical computer and drive the learning device. Meanwhile, $\mathbb{U}^{(k)\dagger}$ is also operated for adjusting its reference axis and prepared for the next iteration. At the agent system, a deterministic single-photon source is fabricated based on the N$_\text{B}$V$_\text{N}$ color-center defect in hexagonal boron nitride (hBN) flakes. More details about the single photon source are shown in the Appendix B. A HWP and PBS initialize the single photons at $|H\rangle$ and a learning device, which is controlled by the computer according to the ``reinforcement'' signal $0/1$ equipped after that. After each iteration, we can use the tomography setup at the agent route to check the fidelity of the state at current step. It is worth to note that during each iteration, only one photon is consumed. In our experiment, we use a chopper (working as a switch) and post-selection method to realize a pseudo-on-demand single photon source~\cite{Yu2018}. That is, when a new iteration begins, we drive the chopper to switch on the register route, and select the first effective coincidence event to be the result~\cite{Yu2018}.

\emph{\bf{Results of the adaptive learning.}}\textemdash For simplicity, we first try to reconstruct a quantum state without imaginary part, e.g., $|E_{1}\rangle=\frac{1}{\sqrt{2}}(|H\rangle+|V\rangle)$, and the experimental results of the semi-quantum reinforcement learning progress are shown in Fig. 3. We can observe that a relative larger reward/punishment ratio $\epsilon$ will cause a slower convergence speed. When $\epsilon=0.80$, the fidelity converges at around $k=36$. However, when $\epsilon=0.50$, the fidelity converges at around $k=6$, which is much earlier than the former case. In this situation, the agent can learn the information of the environment state successfully within 50 iterations. A typical example is shown in Fig. 3, while the fidelity of the agent system can reach over $93\%$ after 20 iterations for all $\epsilon$. We perform 20 identical learning experiments in total for this state and the average fidelity at the last iteration can reach $0.955\pm0.047$, $0.947\pm0.045$, $0.931\pm0.074$ at $\epsilon=0.80, 0.65, 0.50$, respectively~\cite{random,ideale1}.

Furthermore, in order to reveal the benefits or limitations of the our sQRL protocol, we compare it with the standard quantum state tomography (QST) method, which plays as a benchmark here.
We want to note that each iteration just consumes one photonic copy of the environment. Thus, to ensure the QST is performed with the same resources as we used in sQRL, we have to apply only a $k$ number of photons during the QST process, where $k$ denotes the learning step. Since there are in total three measurements that have to be performed in the QST process (more details about the QST process are shown in Appendix C), and each measurement should be provided with the same number of photon copies, we perform the QST every 3 steps. The corresponding results are shown in Fig. 3, denoted by gray diamonds with error bars (the error bars are obtained by 20 independent experiments). The pale blue area shows a convergent period (with respect to $\epsilon=0.5$) in which the fidelities obtained by sQRL results are better than the QST fidelity results. This exhibits that more information of the environment state can be extracted by the sQRL approach, and shows the advantages of sQRL in the case of small number of copies.

\begin{figure}[t]
\centering
\includegraphics[width=0.45\textwidth]{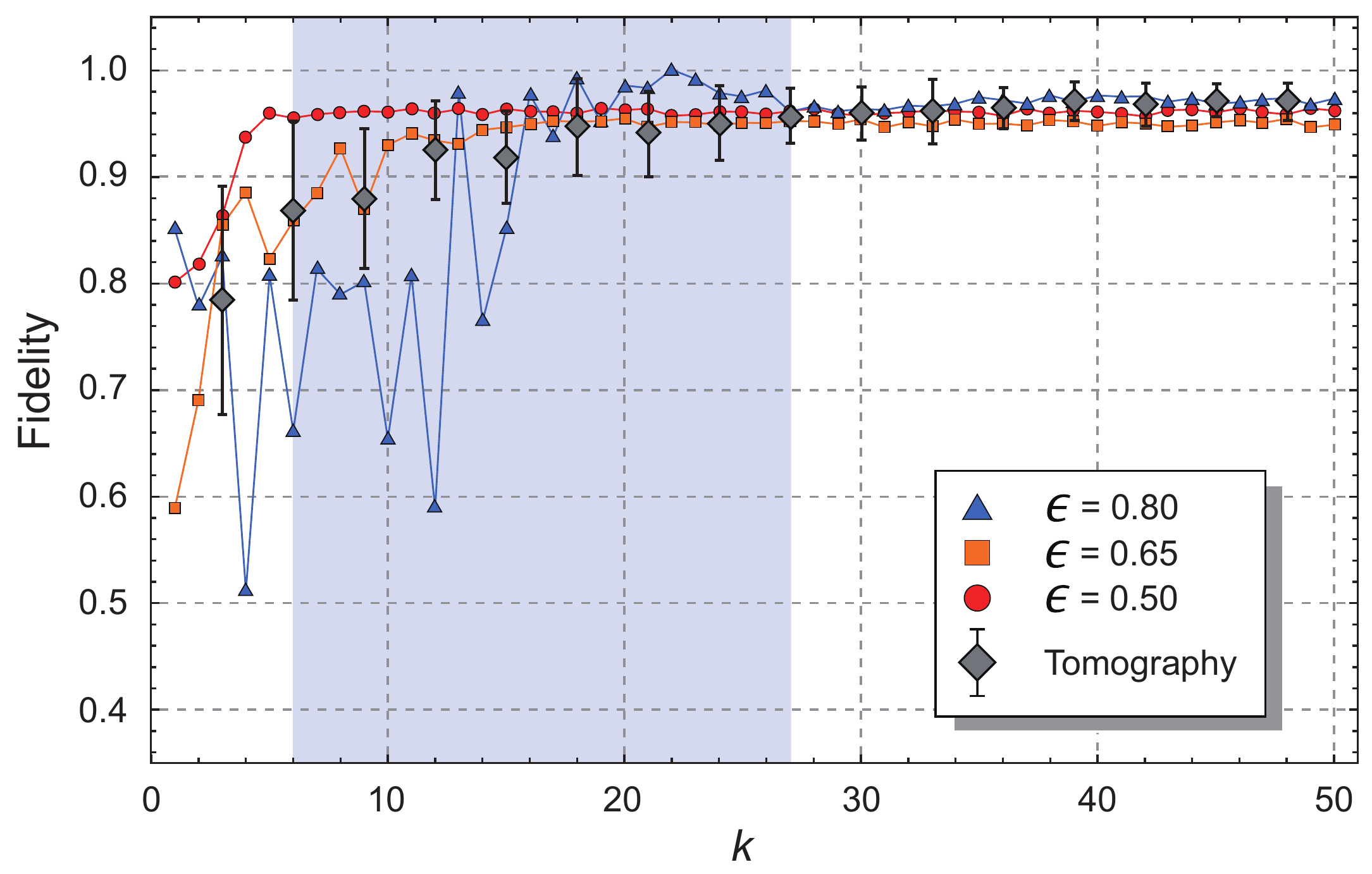}
\caption{Experimental results for the semi-quantum reinforcement learning process with environment state $|E_{1}\rangle$. The triangles (blue), squares (orange), and circles (red) represent the fidelity of the agent state in each iteration with different reward/punishment ratios $\epsilon$. The gray diamonds with error bars denote the average values of fidelities obtained by 20 tomography processes. The pale-blue-shaded area shows the advantages (comparing with the red dots, i.e., choosing $\epsilon=0.5$) of sQRL method which is exhibited in a certain learning step (equal to the photonic copy resource) interval.}
\end{figure}

In order to prove the robustness of our sQRL protocol, we select other two quantum states for which an imaginary part is contained. Figures 4 (a,b) show the learning process for the environment state $|E_{2}\rangle=\frac{1}{\sqrt{2}}(|H\rangle+e^{i\pi/4}|V\rangle)$ and $|E_{3}\rangle=0.948|H\rangle+e^{0.890 i}0.317|V\rangle$, respectively. For these more general environment systems, the sQRL protocol can also successfully adapt the agent state to an ideal fidelity. The average fidelities obtained from 20 learning experiments are $0.886\pm0.033$, $0.882\pm0.035$, $0.860\pm0.060$ at $\epsilon=0.80, 0.65, 0.50$, respectively for $|E_{2}\rangle$; and $0.933\pm0.044$, $0.911\pm0.052$, $0.902\pm0.046$ at $\epsilon=0.80, 0.65, 0.50$, respectively for $|E_{3}\rangle$~\cite{ideale2}. Besides, in these two cases, we find a relatively higher $\epsilon$ (i.e., $\epsilon=0.80, 0.65$) will benefit the agent to learn more information from the environment (i.e., a higher fidelity can be obtained). Still, for these cases, we will compare them with QST results (shown as gray diamonds with error bars in Fig. 4) under the same photonic copy resources. When the environment state is chosen as $|E_{2}\rangle$, we find in all learning conditions (with all three different $\epsilon$), there is not a convergent period in which the fidelities obtained by sQRL can surpass the one get from QST (shown in Fig. 4(b)). This reflects that the ability of the sQRL approach is sensitive to the initial state, namely, for some environment states, the advantage of the sQRL method is not apparent. Besides, we compare the fidelity under state $|E_{3}\rangle$ when $\epsilon$ is set as 0.50 (which begins to converge at $k=6$). We find the fidelity obtained by sQRL in this time can reach to a higher stage until $k=18$ (each basis can be shared six photonic copy resources), which shows that the advantage of the sQRL method is exhibited as well in this occasion.

\begin{figure}[t]
\centering
\includegraphics[width=0.45\textwidth]{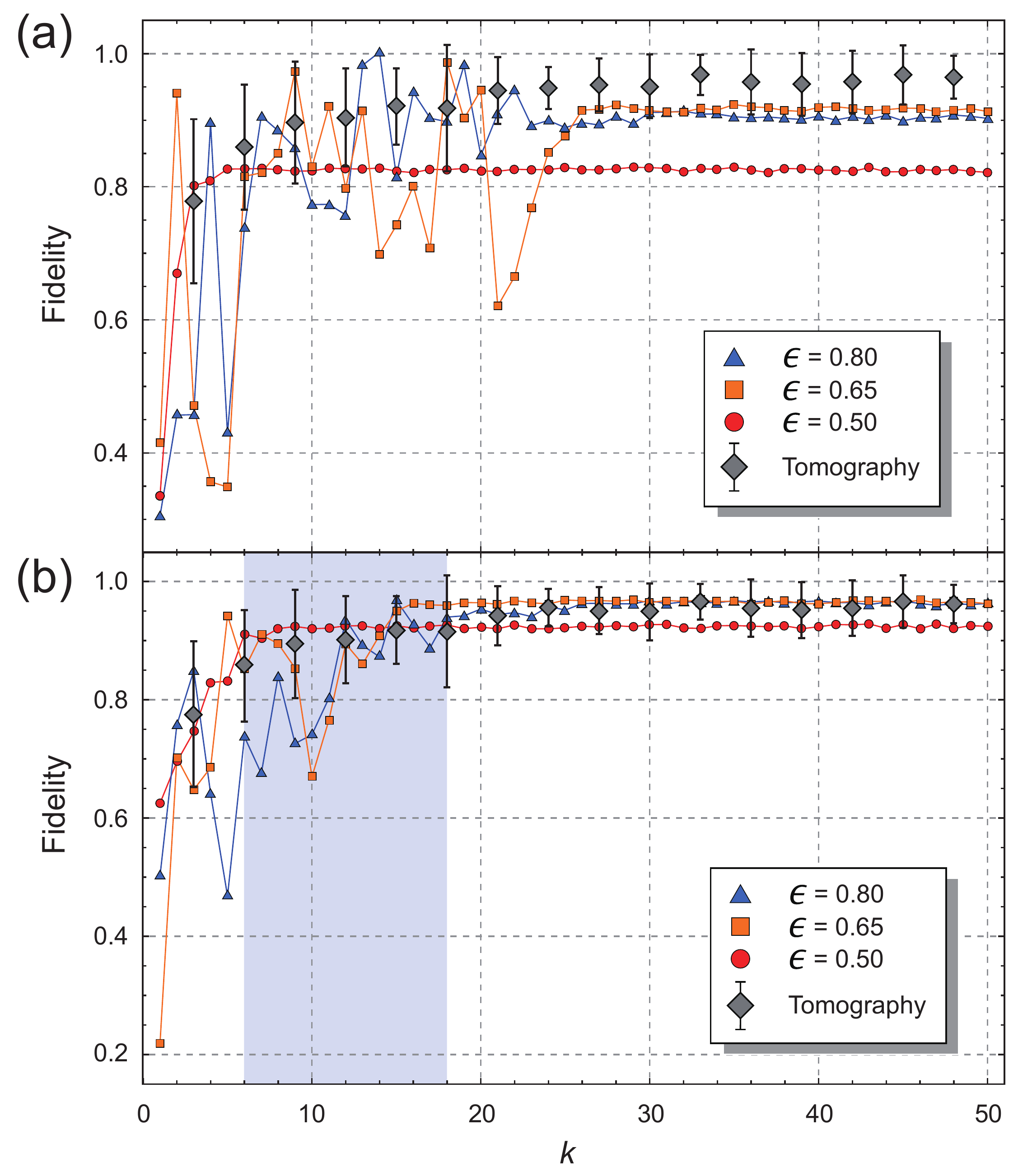}
\caption{(a) The fidelity of the agent state in each step for learning environment $|E_{2}\rangle$. (b) The fidelity of the agent state in each step for learning environment $|E_{3}\rangle$. The symbols are similar to those in Fig. 3.}
\end{figure}

\emph{\bf{Conclusions.}}\textemdash In summary, we demonstrate a semi-quantum reinforcement learning protocol that can make a quantum agent to adapt to an unknown environment qubit state. The environment system can provide a limited number of photonic copies, and the register qubit will interact with it by a CNOT gate. Then, we detect the register photons on $\{|H\rangle,|V\rangle\}$ basis by single-shot measurement, and obtain the ``reinforcement'' signal to decide reward or punishment for the next iteration. Based on the calculation results, we perform the unitary operations on both agent and environment systems. We can observe that, within 50 iterations, the agent can adapt to the three test environment systems with high fidelities under an appropriate reward/punishment ratio $\epsilon$. For example, we have shown cases for test states $|E_{1}\rangle$ and $|E_{2}\rangle$, for which a larger $\epsilon$ produces better results, at the expense of converging later. This is the well-known balance between exploration and exploitation in RL, namely, a larger $\epsilon$ produces larger exploration, with which the final fidelity can be larger, but also more fluctuations, such that the final fidelity is achieved later than with a smaller $\epsilon$. This is a standard characteristic present in reinforcement learning. Furthermore, we compare the sQRL with the QST method. We find that the advantages of the sQRL protocol can be exhibited in many situations, but the protocol is also sensitive to the initial environment state. This sQRL protocol can also be extended to high dimensions (with multilevel gates~\cite{Alber2000}), multipartite (with multiqubit gates~\cite{Zahedinejad2015}), and mixed quantum state situations, which will enable more applications in the quantum information and communication techniques.

\emph{\bf{Appendix A: Realization of CNOT gate.}}\textemdash  Given that in our experiments the control qubit state is arbitrary ($\alpha|H\rangle+\beta|V\rangle$), and the target qubit state is just $|H\rangle$, the CNOT gate is easy to be realized by only one PBS and one HWP. The target qubit state is transformed to be $(|H\rangle+|V\rangle)/\sqrt{2}$ after the HWP @$22.5^{\circ}$. Thus, the whole system can be written as $(\alpha|HH\rangle+\alpha|HV\rangle+\beta|VH\rangle+\beta|VV\rangle)/\sqrt{2}$. We use the post-selection approach here to select the coincidence event of the two outputs, such that the state $(\alpha|HH\rangle+\beta|VV\rangle)/\sqrt{2}$ is obtained, with success probability 1/2~\cite{He2016,Lu2007}.

\emph{\bf{Appendix B: Brief introduction of the hBN material.}}\textemdash The room-temperature single-photon source (SPS) is fabricated based on the hexagonal boron nitride (hBN) flakes performed with the nitrogen-ion irradiation and the high-temperature anneal treatments. The stable single N$_\text{B}$V$_\text{N}$ defects in hBN can be generated and work as SPSs in the existing environment~\cite{Tran2016}. The SPS used in the experiment is excited and collected by the home-made confocal microscope with a NA=0.9 objective. The single-photon purity of the SPS is $g^{(2)}(0)=0.045 \pm 0.045\ll 0.5$, which indicates the remarkable quantum-emission property.

\emph{\bf{Appendix C: Quantum state tomography in experiment.}}\textemdash For the environment state $|E_{1}\rangle$ and choosing the reward/punishment ratio $\epsilon=0.50$, we can find the fidelity begins to converge at $k=6$. This means the agent already obtained the maximal information by consuming only 6 photons. Therefore, in order to guarantee the result obtained from QST method under same resource, we should provide only 6 photons in QST process.

In this situation, we divide the total number of photons (6 photons) into three parts equally, and measure each two photons in the $|H\rangle$/$|V\rangle$ basis, $|H+V\rangle$/$|H-V\rangle$ basis, and $|H+iV\rangle$/$|H-iV\rangle$ basis. Based on the number of the counts in the $|H\rangle$ (or $|V\rangle$), $|H\pm V\rangle$, $|H\pm iV\rangle$ bases, we do a maximum-likelihood estimation to find the closest density matrix and calculate its fidelity~\cite{James2001}.

\section{Acknowledgments}
We acknowledge Nora Tischler for helpful discussions and helping us to make corrections in the early version. This work is supported by the National Key Research and Development Program of China (No. 2017YFA0304100), the National Natural Science Foundation of China (Grants Nos. 61327901, 11674304, 11822408, 61490711, 11774335, and 11821404), the Key Research Program of Frontier Sciences of the Chinese Academy of Sciences (Grant No. QYZDY-SSW-SLH003), the Youth Innovation Promotion Association of Chinese Academy of Sciences (Grants No. 2017492), the Foundation for Scientific Instrument and Equipment Development of Chinese Academy of Sciences (No. YJKYYQ20170032), Anhui Initiative in Quantum Information Technologies (AHY020100, AHY060300), the National Postdoctoral Program for Innovative Talents (Grant No. BX20180293), China Postdoctoral Science Foundation funded project (Grant No. 2018M640587), the Fundamental Research Funds for the Central Universities (No. WK2470000026), Centro Basal FB0807, Ram\'on y Cajal Grant RYC-2012-11391, MINECO/FEDER FIS2015-69983-P, Basque Government IT986-16, and the projects OpenSuperQ (820363) and QMiCS (820505) of the EU Flagship on Quantum Technologies.

\emph{\bf{Conflict of Interest}}

The authors declare no conflict of interest.

\emph{\bf{Keywords}}

reinforcement learning, quantum optics, quantum state reconstruction

\bibliographystyle{}

\begin{thebibliography}{41}
\bibitem{QCRL}
We point out that our semi-quantum reinforcement learning protocol is a quantum-classical type, which combines several iterations of quantum evolution and classical processing. On the one hand, the information that is being learned is quantum, i.e., the environment states, the operations and the measurements are quantum. On the other hand, the feedback mechanism after the measurement is performed classically.
\bibitem{random}
Other nineteen experimental data sets are not shown in the picture.
\bibitem{ideale1}
In the ideal case without experimental noise, we find that the average number of iterations (as for state $|E_{1}\rangle$) to reach the convergence is $k$=17, 9, and 7 for $\epsilon$=0.8, 0.65, and 0.5 respectively. These data are obtained by 20 numerical simulations.
\bibitem{ideale2}
In these two cases (states $|E_{2}\rangle$ and $|E_{3}\rangle$), the average number of iterations to reach the convergence without experimental noise is $k$=27, 11, and 9 for $|E_{2}\rangle$ with $\epsilon$=0.8, 0.65, and 0.5 respectively; and  $k$=26, 18, and 9 for $|E_{3}\rangle$ with $\epsilon$=0.8, 0.65, and 0.5 respectively. These data are obtained by 20 numerical simulations.
\end{thebibliography}

\end{document}